\documentclass[aps,prl,superscriptaddress,showpacs,preprint]{revtex4-1}

\usepackage{amsmath}
\usepackage{amssymb}
\usepackage{amsfonts} 
\usepackage{latexsym}
\usepackage{bbm}
\usepackage{indentfirst} 
\usepackage{graphicx}
\usepackage{subfigure}
\usepackage{mathrsfs}
\usepackage{varioref}

\draft 

\graphicspath{{Figures/}}

\begin{document}


\title{Interaction between electrostatic collisionless shocks generates strong magnetic fields} 



\author{E. Boella}
\email[]{e.boella@lancaster.ac.uk}
\affiliation{Physics Department, Lancaster University, Lancaster, UK}
\affiliation{Cockcroft Institute, Sci-Tech Daresbury, Warrington, UK}
\author{K. Schoeffler}
\affiliation{GoLP/Instituto de Plasmas e Fus\~ao Nuclear, Instituto Superior T\'ecnico, Universidade de Lisboa, Lisbon, Portugal}
\author{N. Shukla}
\affiliation{GoLP/Instituto de Plasmas e Fus\~ao Nuclear, Instituto Superior T\'ecnico, Universidade de Lisboa, Lisbon, Portugal}
\author{G. Lapenta}
\affiliation{Mathematics Department, KU Leuven, Leuven, Belgium}
\author{R. Fonseca}
\affiliation{GoLP/Instituto de Plasmas e Fus\~ao Nuclear, Instituto Superior T\'ecnico, Universidade de Lisboa,  Lisbon, Portugal}
\affiliation{DCTI/ISCTE, Instituto Universitario de Lisboa, Lisbon, Portugal}
\author{L. O. Silva}
\affiliation{GoLP/Instituto de Plasmas e Fus\~ao Nuclear, Instituto Superior T\'ecnico, Universidade de Lisboa, Lisbon, Portugal}

\date{\today}

\begin{abstract}
The head-on collision between electrostatic shocks is studied via multi-dimensional Particle-In-Cell simulations. It is found that the shock velocities drop significantly and a strong magnetic field is generated after the interaction. This transverse magnetic field is due to the Weibel instability caused by pressure anisotropies due to longitudinal electron heating while the shocks approach each other. Finally, it is shown that this phenomenon can be explored in the laboratory with current laser facilities within a significant parameter range.
\end{abstract}

\pacs{}

\maketitle 

Collisionless shocks are ubiquitous in space and astrophysics, where they are considered efficient particle accelerators and they are often invoked to explain, for instance, the cosmic ray spectrum or the emission from gamma-ray bursts \cite{Piran-RMP-2005, Jones-SSR-1191}. During the past few years, the tremendous advances in computer power and numerical modeling have fostered innovative and very detailed kinetic simulations of shocks, allowing for the comprehension of many aspects of the shock micro-physics, from shock formation to particle acceleration \cite{Ruyer-PRL-2017,Bret-POP-2014,Bret-POP-2013,Martins-APJL-2009, Spitkovsky-APJL-2008a, Spitkovsky-APJL-2008b}. However, to date almost no theoretical, computational, or experimental studies investigating the interaction among collisionless shocks are available. The phenomenon is equally important and quite pervasive in astrophysics: the internal shock model of GRB is perhaps the most well-known example where shock collision is thought to play a role \cite{Rees-RAS-1978}. Furthermore, very recently shock collision has been observed for the first time in the extragalactic jet of the nearby radio galaxy 3C 264 \cite{Meyer-NAT-2015}.

The strong advance of laser technology motivated the first experimental studies of collisionless shock generation in laboratory plasmas \cite{Palmer-PRL-2011, Haberberger-NP-2012, Tresca-PRL-2015, Antici-exploded, Chen-Marija,  Albert-APS-2017}. In particular, the pioneering demonstration of laser-driven electrostatic shocks in near-critical density plasmas has received great prominence, due to its potential for producing mono-energetic ion beams \cite{Haberberger-NP-2012, Fiuza-PRL-2012, Fiuza-PoP-2013, Boella-buh}. While the conditions for generating full relativistic electromagnetic shocks of the type more common in astrophysical settings in the laboratory are not completely understood \cite{ Park-POP-2015, Park-Conf-2016,Ross-PRL-2017}, with this article, we propose to take advantage of the solid experimental results on electrostatic shocks obtained so far to explore the interaction among these collisionless shocks in the laboratory. A novel configuration to investigate the binary collision of shock waves exploiting near critical density target(s) heated by laser beams is suggested. Performing \textit{ab initio} multi-dimensional Particle-In-Cell (PIC) simulations, we investigate in detail the microphysics of the interaction and identify a regime for shock collisions, where kinetic effects determine the interaction dynamics. The results indicate that the collision does not change the nature of the waves, with the shocks being able to pass through each other and proceed in opposite directions, similarly to what predicted for electrostatic solitons \cite{Zabusky-PRL-1965}. However, the collision is highly inelastic causing the shocks to slow down considerably after the interaction. A strong perpendicular magnetic field develops right after the collision. The field is generated by the Weibel instability \cite{Weibel-PRL-1959,Bret-APJ-2009,Bret-POP-2010,Shukla-JPP-2012,Shukla-JPP-2018} which is driven by an electron pressure anisotropy in the interaction region, caused by strong longitudinal heating occurring while the shocks are approaching. Finally, estimates of the laser parameters and target density which will allow for generating counter-propagating shocks in the laboratory and for the magnetic field to be observed after their collision, are provided. 

\begin{figure*}[]
\centering
\includegraphics[scale=0.17]{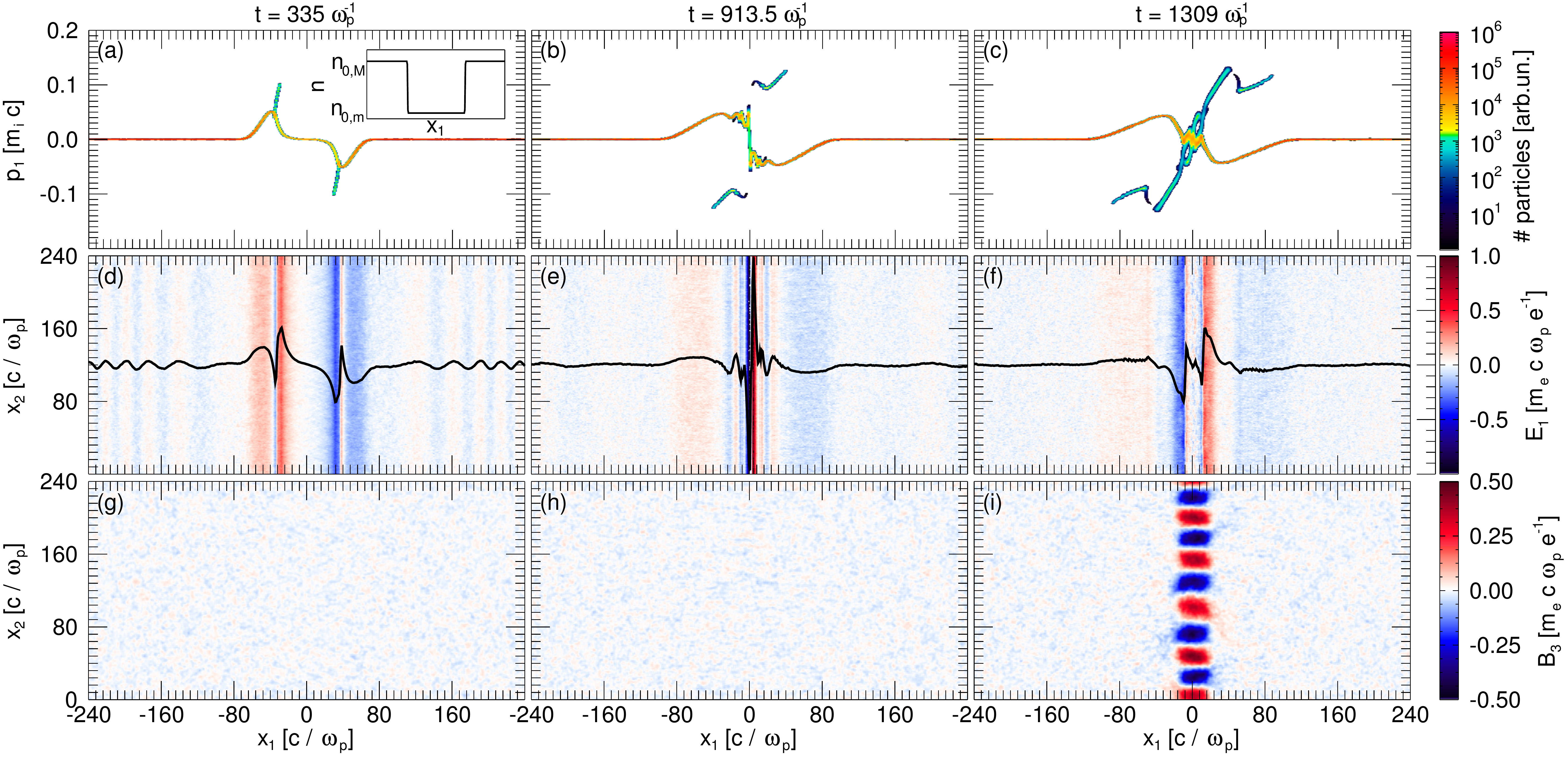}
\vspace{-6pt}
\caption{Longitudinal ion phase space (a-c), longitudinal electric field $E_1$ (d-f) and transverse magnetic field $B_3$ (g-i) at shock formation ($t = 335 \, \omega_p^{-1}$, first column), shock collision ($t = 913.5 \,  \omega_p^{-1}$, second column) and after the interaction ($t = 1309 \,  \omega_p^{-1}$, third column). The inset in (a) shows the initial density profile with the sharp discontinuities which trigger the formation of the two counter-propagating shocks. The black solid lines in (d-f) are an average of $E_1$ along the $x_2$ direction.} \label{Fig1}
\end{figure*}

In order to study the interaction of two collisionless electrostatic shocks, we performed a series of one- and two-dimensional (1D and 2D, respectively) simulations using the PIC code OSIRIS \cite{Fonseca-LNCS-2002}. A plasma composed by uniformly hot Maxwellian electrons and cold hydrogen ions with realistic charge-to-mass ratio (akin to a laser-generated system) is considered. The two species do not have any initial drift velocity (this is a valid assumptions when considering laser pulses interacting with near critical density targets, where most of the laser energy will be transferred to the plasma in the form of electron heating \cite{Fiuza-PRL-2012}). The plasma density profile presents two perfectly symmetric sharp discontinuities (see inset in Fig. \ref{Fig1} (a)). They constitute the jump in density that triggers the shock formation, similarly to that explained in \cite{Fiuza-PoP-2013, Dieckmann-PPCF-2010}. We have carried out simulations, considering electron temperatures in the range $T_e = 0.2 \cdot 10^{-3} \, - \, 1.5 \, \text{MeV}$ and initial density jumps in the interval $\sigma \equiv n_{0,M}/n_{0,m} = 3.33 \, - \, 10$, where $M$ and $m$ indicate the highest and the lowest density respectively. The largest simulation window adopted was $4800 \, c/\omega_p$ long and $240 \, c/\omega_p$ wide, with $c/\omega_p$ the electron skin depth, $c$ the speed of light in vacuum, $\omega_p \equiv \sqrt{4 \pi e^2 n_{0,M}/m_e}$ the electron plasma frequency, $e$ the elementary charge and $m_e$ the electron mass. The system is numerically resolved with 4 cells per skin depth or electron Debye length $\lambda_D \equiv \sqrt{k_B T_e/4 \pi e^2 n_{0,M}}$, where $k_B$ is the Boltzmann constant, and the temporal step is chosen to satisfy the Courant condition. In order to model the plasma dynamics correctly 36 (or 1000 in 1D) particles per cell and quartic interpolation were employed. Periodic boundary conditions have been used, but the simulation window is large enough that they do not interfere with the plasma dynamics.

In Fig. \ref{Fig1}, the main stages of a typical simulation are reported. In this case, $T_e = 1.5 \, \text{MeV}$ and $\sigma = 10$ were used. When the more dense plasma expands into the less dense component, it drives several non linear-waves that eventually develop into two shock waves streaming towards each other in opposite directions. The shock structures are clearly recognizable at $x_1 \simeq  \pm 35 \, c/\omega_p$ in Figs. \ref{Fig1} (a) and (d), which depict the longitudinal ion phase space and the longitudinal electric field $E_1$ at $t=335 \, \omega_p^{-1}$. In particular, $E_1$ presents a double layer structure, which is a typical signature of electrostatic shocks \cite{Sagdeev-RPP-1966, Sorasio-PRL-2006, Stockem-PRE-2013}. The shock waves collide in the middle of the simulation window ($x_1 = 0$) at $t = 913.5 \, \omega_p^{-1}$ (Figs. \ref{Fig1} (b) and (e)). The collision results in an enhancement of the electric field (Fig. \ref{Fig1} (e)). It is interesting to observe that until $t = 913.5 \, \omega_p^{-1}$, the magnetic field in the direction out of the box $B_3$ is negligible (Figs. \ref{Fig1} (g) and (h)). After the collision, the shocks are able to pass through each other and continue streaming in opposite directions (Figs. \ref{Fig1} (c) and (f)). However $B_3$ in the middle of the simulation window presents an intense filamentary structure, which has developed only after the interaction. The field is confined in a small region of the simulation window, between $x_1 = [-30, \, 30] \, c/\omega_p$ corresponding to the downstream of the shocks. The filaments are characterized by a wavelength $\lambda \simeq 50.7 \, c/\omega_p$. The field reaches a maximum at around $t \simeq 1309 \, \omega_p^{-1}$ and then starts to decay.
\begin{figure}[]
\centering
\includegraphics[scale=0.17]{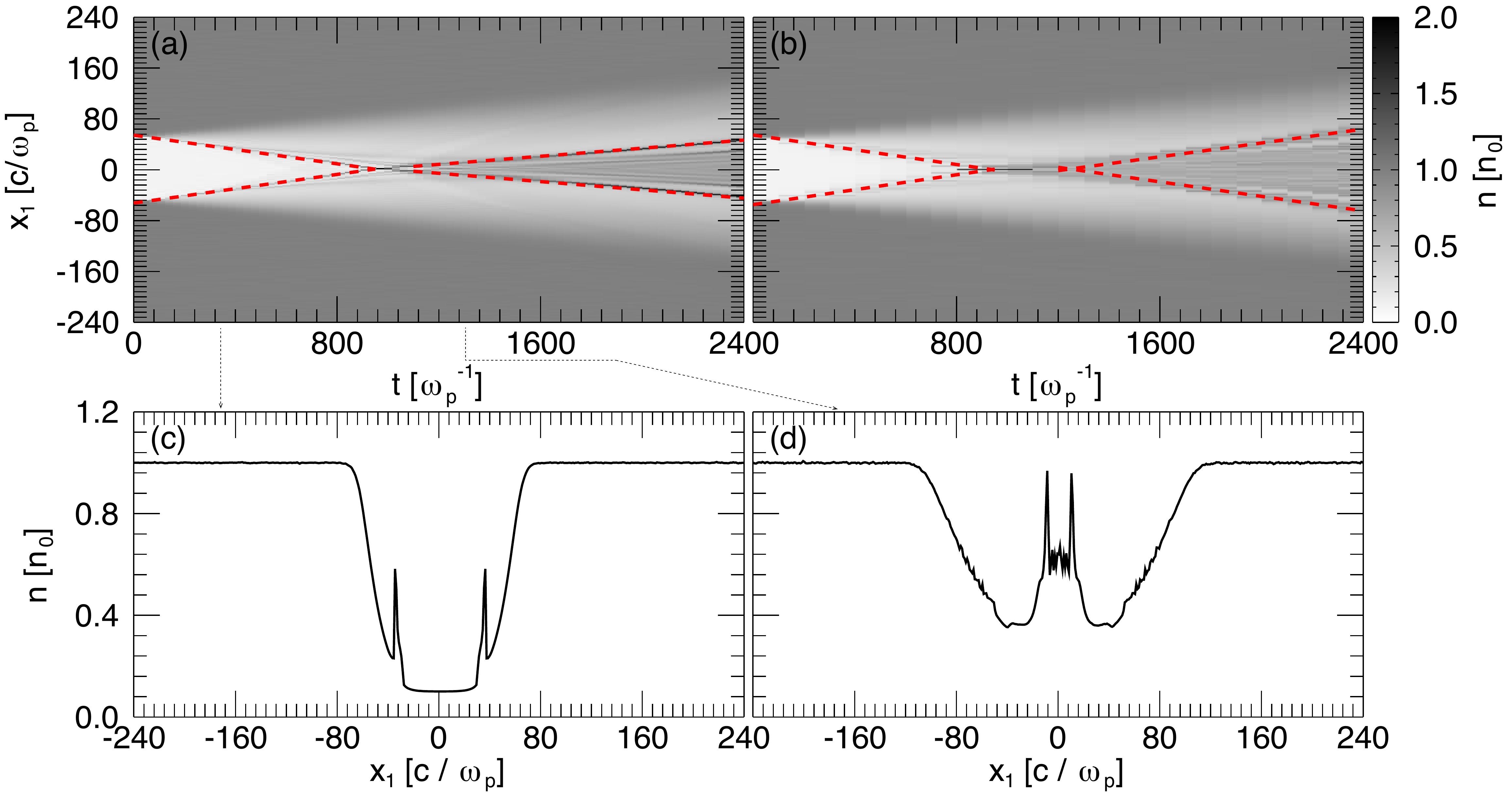}
\vspace{-6pt}
\caption{Evolution of the ion density for a 2D (a) and a 1D simulation (b). The values in (a) result from an average along the $x_2$ direction. The transversely averaged density is shown in detail at shock formation ($t=335 \, \omega_p^{-1}$, Fig. (c)) and after the collision ($t=1309 \, \omega_p^{-1}$, Fig. (d)). The red dashed lines follow the shocks and their slope measures the shock speed, which is $v_s=\pm 0.056 \, c$ and $\pm 0.032 \, c$ before and after the collision respectively in (a) and $v_s=\pm 0.058 \, c$ and $\pm 0.054 \, c$ before and after the collision respectively in (b).} \label{Fig2}
\end{figure}

We measured the velocity of the shocks before and after the interaction. Results are shown in Fig. \ref{Fig2} (a), where the evolution of the ion density averaged along the $x_2$ direction is plotted. The shocks move at a constant speed $v_s=\pm 0.056 \, c$ corresponding to a Mach number $\mathcal{M}=v_s/c_s = 1.4$, where $c_s=\sqrt{k_BT_e/m_i}$ is the sound speed for ions of mass $m_i$, in excellent agreement with the theoretical model in \cite{Stockem-PRL-2014}. After the collision, the speed of both shocks drops to $v_s \simeq \pm 0.032 \, c$, which corresponds to about 42\% of its initial value. We notice that, despite the different simulation setup, the slow-down agrees moderately well with the estimates in \cite{Nitin-DM}, which predicts a decrease in velocity $\approx 1/\sqrt{2}$  and $\approx 1/\sqrt{3}$ in 2D and 3D simulations of interpenetrating plasmas respectively. The reasons for this will be clear after the analysis of the particle distribution functions, presented in the next paragraphs. For comparison, we report the same results as obtained from a 1D simulation (Fig. \ref{Fig2} (b)), where, due to the geometry, the growth of modes perpendicular to $x_1$ is inhibited and therefore no $B_3$ is observed. In this case the shocks only slightly slow down: their velocity passes from $v_s = \pm 0.058 \, c$ to $v_s \simeq \pm 0.054 \, c$ after the collision. We verified that despite the slowdown, the nature of the shock waves remain unchanged: the electric field maintains the double layer structure (Figs. \ref{Fig1} (d) and (f)), the density jump conditions for low Mach number shocks \cite{Forslund-PRL-1971} are satisfied (Figs. \ref{Fig2} (c) and (d)) and the waves are moving with $\mathcal{M} > 1$. These results indicate that the multidimensional interaction of collisionless shocks is an inelastic process and that the waves lose energy during the collision much more substantially in 2D than 1D. Given this and the prediction in \cite{Nitin-DM}, we expect that in 3D simulations the shock velocity after the collision would be even lower.
\begin{figure*}[]
\centering
\includegraphics[scale=0.17]{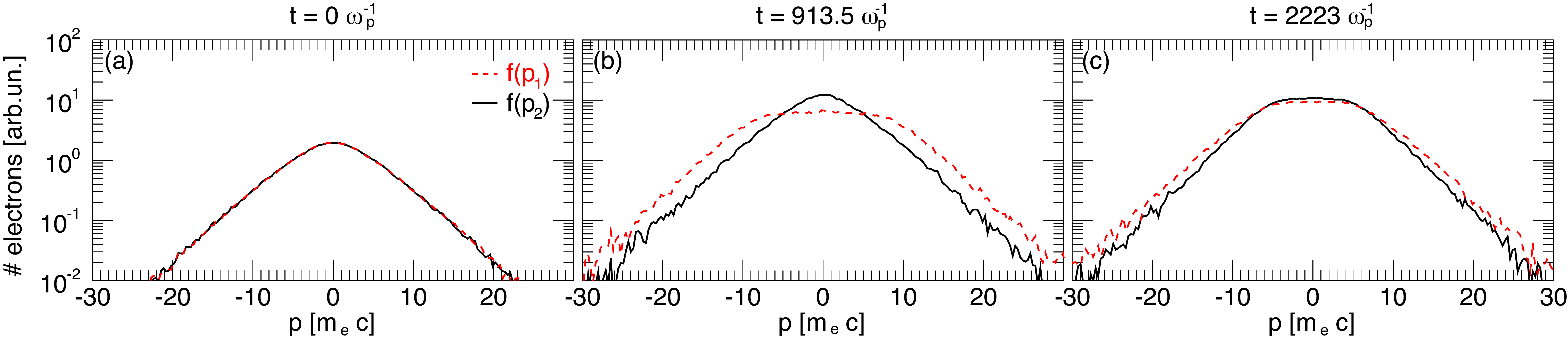}
\vspace{-6pt}
\caption{Electron distribution functions $f(p_1)$ (red dashed) and $f(p_2)$ (black solid) in the middle of the simulation window ($x_1 = 0$) at the beginning of the simulation (a), collision time (b) and much after the interaction (c).} \label{Fig2_bis}
\end{figure*}

To gain insight over what happens immediately after the collision, we have checked the electron distribution functions $f(p_1)$ and $f(p_2)$, with $p_1$ and $p_2$ longitudinal and traverse momentum respectively, in the interaction region (Fig. \ref{Fig2_bis}) and far from it (not presented here) much before, during and after the encounter. At the beginning of the simulation, $f(p_1)$ and $f(p_2)$ are perfectly Maxwellian everywhere in the simulation window (Fig. \ref{Fig2_bis} (a)). At $t = 913.5 \, \omega_p^{-1}$ and around $x_1 = 0$,  $f(p_1)$ is much broader than $f(p_2)$ (Fig. \ref{Fig2_bis} (b)). The longitudinal field $E_1$, which is far from being uniform in the region corresponding to the upstream of the two shocks, slightly before the collision, heavily heats the electrons in the longitudinal direction. This does not happen in the downstream of the shocks, where $f(p_1)$ and $f(p_2)$ remain equivalent. We observe that both our simulation setup and our results are fundamentally different from \cite{Stockem-PRL-2014}. In \cite{Stockem-PRL-2014}, strong longitudinal electron heating leading to Weibel generated fields was detected in the downstream of an electrostatic shock driven by the interpenetration of plasma shells moving towards each other with an equal and opposite drift velocity; simulations of single shocks triggered by a density jump in plasmas with no drift show no magnetic field evidence in the downstream region and lead to constant shock velocities \cite{Fiuza-PoP-2013, Boella-buh}. Further in time, the distribution functions in what is now the downstream of the shocks become very similar, but their spread is bigger than in the far upstream (Fig. \ref{Fig2_bis} (c)).

\begin{figure}[]
\centering
\includegraphics[scale=0.17]{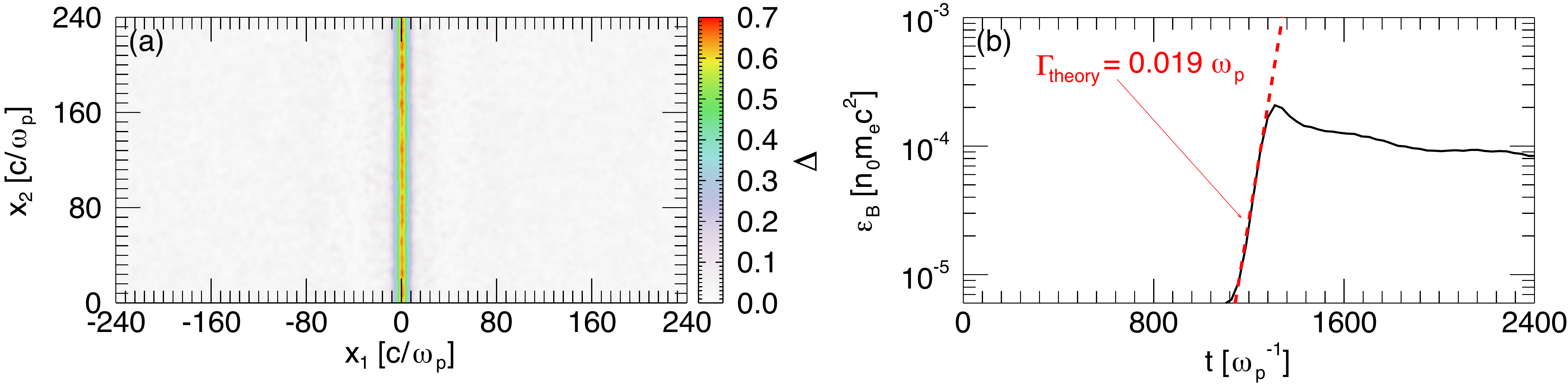}
\vspace{-6pt}
\caption{Anisotropy $\Delta$ at $t = 974.4 \, \omega_p^{-1}$ when it reaches the maximum value $\Delta_{\text{max}}=0.7$ (a) and magnetic field energy density in the region between $x_1 = -30 \, c/\omega_p$ and $x_1 = 30 \, c/\omega_p$ (b). The superimposed red dashed line indicates the theoretical growth rate $\Gamma_{\text{theory}}=0.019 \, \omega_p$.} \label{Fig3}
\end{figure}

The electron isotropization could then explain the agreement between the slowdown of the shock waves and \cite{Nitin-DM}. In fact, in the reference frame of the shock, the electrons are moving towards the shock with longitudinal velocity $-v_s$ and the magnetic field $B_3$ bends them in the $x_2$ direction, thus equally equipartitioning their kinetic energy among the two in-plane directions. As a result, the longitudinal component of their velocity decreases as $1/\sqrt{2}$, which translates in the same shock slow-down when moving back to the simulation reference frame (e.g. downstream reference frame). We computed the anisotropy $\Delta = \alpha_{\parallel}/\alpha_{\perp}-1$, with $\alpha_{\parallel}$ and $\alpha_{\perp}$ defined according to \cite{Yoon-PoP-2007}:
\begin{eqnarray}
\frac{T_{e\perp}}{m_e c^2} &=& \frac{1}{\alpha_{\perp}}, \\
\frac{T_{e\parallel}}{m_e c^2} &=& \frac{1}{\alpha_{\parallel}}\left [ 1+\left(\frac{\alpha_{\parallel}}{\alpha_{\perp}}\right) \frac{K_1\left(\alpha_{\parallel}\right)}{\alpha_{\parallel} K_2\left(\alpha_{\parallel}\right)} \right]^{-1},
\end{eqnarray}
where $K_n$ indicates the modified Bessel function of the second kind of order $n$ and  $T_{e\perp}$ and $T_{e\parallel}$ are calculated from the temperature tensor $\mathbb{T}$ performing a matrix diagonalization as described in \cite{Kevin-diagonalization}. The result of the latter operation is such that $\mathbb{T}_{1,1}  = T_{e\perp}$ and $\mathbb{T}_{2,2}=T_{e\parallel}$, where the symbols $\perp$ and $\parallel$ refer to perpendicular and parallel respect to the wave vector of the instability and identify the direction with the higher and lower temperature, respectively.  Figure \ref{Fig3} (a) reports $\Delta$ in the simulation window at $t=974.4 \, \omega_p^{-1}$, when the anisotropy reaches its maximum $\Delta_{\text{max}}=0.7$. It is interesting to observe that, apart from the central region, the anisotropy in the rest of the domain is almost zero. Given $\alpha_\perp \simeq 0.19$, $\Delta = 0.7$ and the wavenumber $k = 2 \pi/\lambda=0.12 \, \omega_p/c$ as provided by the simulation and employing the formula in  \cite{Yoon-PoP-2007}, we have computed the growth rate of the electron Weibel instability $\Gamma_{\text{theory}}=0.019 \, \omega_p$. We have compared the theoretical prediction with the numerical results. Figure \ref{Fig3} (b) shows the evolution of the energy of $B_3$ in the region between $x_1 = -30 \, c/\omega_p$ and $x_1 = 30 \, c/\omega_p$. The magnetic field energy has an exponential growth until $t \simeq 1309 \, \omega_p^{-1}$, where it reaches a maximum and then it starts decaying. The slope of the linear phase allows for inferring the growth rate of the instability, which is measured to be $0.02 \, \omega_p$ in excellent agreement with the theory. This is a strong evidence that the Weibel instability is at play and causes the magnetic field to grow, leading to the isotropization of the electrons. However, we notice that the simulation wavenumber and the relative growth rate do not correspond to the fastest growing mode. We speculate that this is probably due to the fact that the latter rises within a very short period of time and saturates quickly; thus the only modes surviving and observable are those at lower $k$s. The evolution of the Fourier transform of $B_3^2$ (not shown here) displays indeed low energetic modes at $k \simeq 0.25 \,  \omega_p/c$, which saturate in a time interval of $\approx 100 \, \omega_p^{-1}$ after the collision.

It is thus the Weibel instability that causes the shocks to slow down. The intense longitudinal electron heating occurring while the shocks are approaching produces a temperature anisotropy, which in turns drives the electron Weibel instability in the interaction region. The instability saturates when the distribution functions are isotropic again. In the 1D case, the Weibel instability, whose modes are perpendicular to $x_1$, cannot occur and therefore the shock velocity is only slightly decreased by the longitudinal heating.

\begin{figure}[b]
\centering
\includegraphics[scale=0.17]{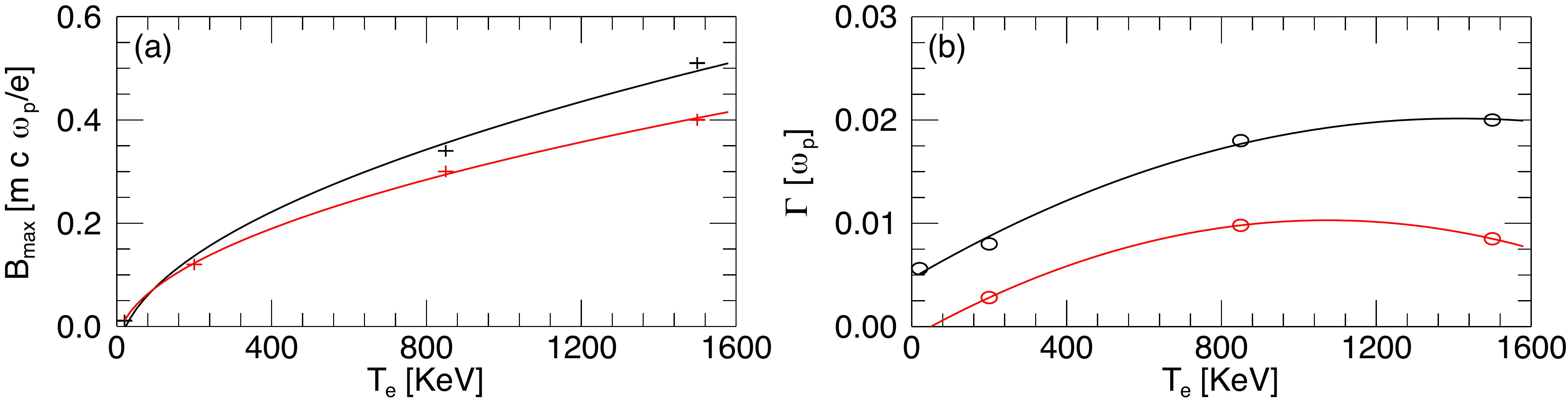}
\vspace{-6pt}
\caption{Maximum amplitude of the magnetic field (a) and instability growth rate (b) versus $T_e$ for $\sigma = 10$ (black) and $\sigma = 5$ (red) as given by the simulations. The magnetic field amplitude scales as $\sqrt{T_e}$, while the data for the growth rate are fitted with quadratic curves.} \label{Fig4}
\end{figure}
The ability to drive two counter-propagating shocks experimentally, whose collision will lead to the growth of the Weibel instability, is connected to the laser intensity and the target maximum density and thickness. In fact, these parameters affect the initial density jump $\sigma$ and electron temperature $T_e$, which trigger the shock formation and determine the shock initial speed, thus influencing the instability to be observed. To understand the dependence of the maximum amplitude of the magnetic field and the growth rate of the instability, both measurable in experiments, on $T_e$  and $\sigma$, we have performed a parameter scan varying these quantities (Fig. \ref{Fig4}). The field amplitude at saturation varies as $\sqrt{T_e}$ (Fig. \ref{Fig4} (a)): saturation of the field is expected when the magnetic pressure $\propto B^2$ equals the plasma pressure $\propto T_e$. The growth rate values are fitted by quadratic curves (Fig. \ref{Fig4} (b)); the quadratic trend can be explained by the larger electron inertia of more energetic electrons. Given the higher inertia, they are more difficult to isotropize and the time required for the Lorentz force to accomplish this, therefore, longer. It is then expected that for even higher temperature, not of interest for this context, $\Gamma$ will decrease with increasing $T_e$. In order to compare our findings with current laser experiments, we compute the laser normalized intensity using the ponderomotive scaling formula \cite{Wilks-PRL-1982}. To reach the considered electron temperatures, laser pulses with normalized vector potential $a_0 = \left(k_B T_e/m_e c^2 + 1 \right)^2-1 \simeq 1-14$ are necessary. We note that such intensities are already or will be soon available, creating an opportunity to probe the setup here proposed. The other crucial factor to drive an electrostatic shock is the formation of a sharp density variation in the plasma \cite{Fiuza-PRL-2012}. Sharp gradients comparable to those employed in our simulations can be easily obtained in the laboratory using thin targets whose density profile reaches a maximum close to the critical density \cite{Fiuza-PoP-2013, Tresca-PRL-2015, Albert-APS-2017}. In fact, the laser will be stopped around the critical density and its radiation pressure will contribute to further steepen the density, which will lead to the formation of the shock. 

In summary, we have performed fully kinetic simulations of the interaction between two electrostatic shocks. The resulting head-on collision was highly inelastic, with the shocks slowing down to up to 50\% with respect to their initial velocity. The decrease in shock velocity is due to a strong magnetic field in the direction perpendicular to the simulation plane produced by the Weibel instability. The temperature anisotropy, which drives the electron Weibel instability after the collision, is caused by an intense longitudinal electron heating occurring upstream of the shocks, as they approach each other. The magnetic field starts decaying when the electron distributions are again isotropic. We have confirmed the magnetic field growth due to the Weibel instability for different initial plasma conditions attainable with available or near future laser beams, thus suggesting that the physics of shock collision can currently be investigated in the laboratory. For instance, the $10 \, \mu \text{m}$ CO$_2$ laser systems available at University of California at Los Angeles \cite{Haberberger-OE-2010} and at the Accelerator Test Facility at the Brookhaven National Laboratory \cite{Polyanskiy-OE-2011} with $a_0$ in the range $1.5-2.5$ would easily allow for testing this setup with commercial hydrogen gas jets. In principle the scheme could be tested also with near-infrared lasers. However in this case targets with higher densities are required \cite{Flacco-NP-2015}. Finally, we note that the proposed setup allows for probing the electron Weibel instability in a fully collisionless regime in the laboratory, similarly to what has previously been done for the ion Weibel instability \cite{Fox-PRL-2013, Huntington-NP-2015}.


%
%

%

\begin{acknowledgments}
This work was supported by the European Research Council (ERC-2015-AdG Grant No. 695008), the Onderzoekfonds KU Leuven (Research Fund KU Leuven, GOA scheme and Space Weaves RUN project) and the US Air Force EOARD Project (FA2386-14-1-0052). Simulations were performed at the Accelerates cluster (Lisbon, Portugal) and on the supercomputer Marconi (Cineca, Italy) under PRACE allocations. 
\end{acknowledgments}

\bibliography{shock_collision}

\end{document}